\documentclass[aps,pra,twocolumn,letterpaper,groupedaddress,10pt]{revtex4-2}
\usepackage{amssymb,amsthm,amsmath,amsfonts}
\usepackage{graphicx,ulem,enumerate,bbm,bm}
\usepackage[pdftex,dvipsnames,usenames]{xcolor}
\usepackage[colorlinks=true,urlcolor=blue,citecolor=blue,linkcolor=blue]{hyperref}

\begin{document}

\title{Quantum walks and entanglement in cavity networks}

\author{Christian Di Fidio}
    \affiliation{Paderborn University, Institute for Photonic Quantum Systems (PhoQS), Theoretical Quantum Science, Warburger Stra\ss{}e 100, 33098 Paderborn, Germany}

\author{Laura Ares} 
    \email{laurares@mail.uni-paderborn.de}
    \affiliation{Paderborn University, Institute for Photonic Quantum Systems (PhoQS), Theoretical Quantum Science, Warburger Stra\ss{}e 100, 33098 Paderborn, Germany}

\author{Jan Sperling} 
    \affiliation{Paderborn University, Institute for Photonic Quantum Systems (PhoQS), Theoretical Quantum Science, Warburger Stra\ss{}e 100, 33098 Paderborn, Germany}

\date{\today}

\begin{abstract}
    For harnessing the full potential of quantum phenomena, light-matter interfaces and complexly connected quantum networks are required, relying on the joint quantum operation of different physical platforms.
    In this work, we analyze the quantum properties of multipartite quantum systems, consisting of an arbitrarily large collection of optical cavities with two-level atoms.
    In particular, we explore quantum walks in such systems and determine the resulting entanglement.
    Realistic imperfections are included in the model as optical losses and spontaneous decays of atoms.
    The topology of torus and the nonorientable M\"obius strip serve as examples of complex networks that we consider, demonstrating the versatility of our approach and resulting in interesting quantum dynamics and interference effects for quantum simulation applications.
\end{abstract}

\maketitle

\section{Introduction}
\label{sec:Introduction}

    Entanglement has attracted a lot of attention since the early days of quantum mechanics \cite{S35.1,S35.2,S36} as it presents a fundamental physical concept \cite{EPR,Bell,Bell1}.
    Nowadays, quantum correlations receive ever increasing recognition within the rapidly developing fields of quantum technology, quantum information, and quantum computation;
    see, e.g., Refs. \cite{Nielsen,Haroche_1,Wilde} for overviews.
    In particular, quantum entanglement is considered to be the characteristic feature that allows one to share quantum information beyond classical limitations, and it provides a central resource for realizing quantum communication protocols \cite{Nielsen1,Horodecki1,Guehne,Pan,Flamini}.

    One prominent quantum application is the quantum simulator that is able to simulate any quantum process \cite{Feynman,Childs,Lovett}.
    As random walks can mimic any classical stochastic process, their quantum counterpart, named quantum walks, can realize the dynamics of arbitrary quantum systems, thus satisfying the defining features of a quantum simulator \cite{Aharonov,Kempe,Kempe1,Andraca,Ambainis,Portugal}.
    Since quantum walks concern the quantum evolution of large composite systems, a natural connection between this concept and the notion of highly multipartite entanglement exists. 

    Because of the aforementioned relevance, quantum walks have been implemented in a large variety of physical systems.
    In quantum optics, for example, this is usually achieved via various linear optical networks \cite{Bouwmeester,Manouchehri,Knight,Hillery,Jeong,Do,Francisco}.
    For instance, input nonclassicality is converted to output entanglement in such scenarios via simple beam splitters \cite{Kim,Xiang,Vogel}.
    Other optical schemes are based, for example, on photonic lattices \cite{Duer,Rai,Schreiber1,Heckelmann,Zurita}.
    Moreover, quantum simulators with a nonlinear quantum-walk evolution have been studied \cite{Peruzzo,Schreiber2,Geraldi,Vakulchyk,Benlloch,Agarwal,Held}, and imperfect quantum walks on graphs with decoherence have been investigated \cite{Paris1,Paris2}.
    Furthermore, quantum walks have been realized with single optically trapped atoms \cite{Karski} and in ion traps \cite{Schmitz,Zaehringer}.
    \textcolor{black}{Another promising quantum-walk platform is optomechanical and electromechanical systems \cite{Aspelmeyer,Oliveira,Mao}.}

    In general, trapped atoms and ions interacting with quantized light fields are well-suited candidates for the realization of hybrid quantum information processing \cite{Cirac1,Monroe,Mabuchi,Srinivas}.
    The combination of long-lived atomic states and light fields can be highly beneficial in quantum networking and for distributed quantum computation \cite{Cirac,Milburn,Weinbrenner}.
    Moreover, various achievements in cavity QED and in tapped-ion techniques have rendered it possible to experimentally generate pairs of entangled atoms \cite{Haroche_2}, to create entangled states of several atoms \cite{Wineland}, and to maintain robust entanglement of macroscopic ensembles of atoms \cite{Polzik}.
    Light-matter interfaces also naturally provide a link between the continuous-variable entanglement of photons \cite{Silberhorn} and cold atoms \cite{Josse}.

    In the present contribution, we combine entanglement theory, quantum simulation, and cavity QED to study the generation and distribution of entanglement in networks.
    In view of the widespread applications of cavity-assisted single-photon sources, we analyze composite quantum systems consisting of atom-cavity nodes that are connected in a cascaded configuration \cite{Carmichael_1,Gardiner_1,Carmichael_2}.
    As found for smaller cascades \cite{Parkins,Carmichael_3,Cirac_1}, one of the advantages of using such configurations is that the open-system evolution itself creates the entanglement \cite{DiFidio_1,DiFidio_2,Wang,Moelmer_1}, even extending to other scenarios, such as optomechanical systems \cite{Rafeie}.
    Here, we consider an arbitrary network of cavities that are optically coupled, including realistic imperfections.
    Initializing the system with a single excitation, we explore how this excitation propagates between quantized radiation fields and the matter and is simultaneously spread across the network.
    This implements a quantum walk that can be compared with the analogous classical random walk.
    Moreover, we analyze the multipartite entanglement dynamics that is established in the networks between the different nodes, as well as the light-matter entanglement, by applying tailored multipartite entanglement witnesses.
    Elaborate network configurations as nontrivial graphs are characterized.

    The paper is organized as follows.
    In Sec. \ref{sec:LightMatterDynamics}, the master equation describing the dynamics of a single atom-cavity system is reviewed.
    In Sec. \ref{sec:CascadedNetworks}, the master equation describing the dynamics of an arbitrary multi-cavity system is formulated, and the evolution is solved analytically by means of the quantum trajectory method.
    In Sec. \ref{sec:Geometries}, we study different network geometries, specifically a network with a toroidal structure and one with a M\"obius-strip configuration.
    We further analyze how an initial excitation propagates, providing a light-matter-interfacing  quantum walk.
    In Sec. \ref{sec:EntanglementEvolution}, the entanglement between the different constituents of the full quantum system is analyzed, including the impact of attenuations.
    Finally, concluding remarks are offered in Sec. \ref{sec:Conclusion}.

\section{Damped Light-Matter Dynamics}
\label{sec:LightMatterDynamics}

\begin{figure}[b]
    \centering
    \includegraphics[width=0.8\columnwidth]{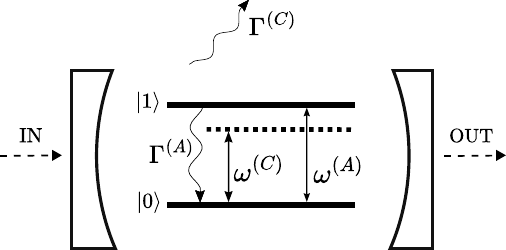}
    \caption{%
        Elementary building block of the cascaded network.
        A cavity mode of frequency $\omega^{(C)}$ is detuned from a two-level atomic transition of frequency $\omega^{(A)}$.
        All optical losses are collected in the decay rate $\Gamma^{(C)}$, and $\Gamma^{(A)}$ is the spontaneous emission rate of the atom, $|1\rangle\to|0\rangle$.
    }\label{fig:LeakyCavity}
\end{figure}

    Before considering an entire network, we analyze a single building block that is used in the next section as a node to construct the cascaded network.
    Specifically, we consider a high-$Q$ optical cavity, supporting one monochromatic mode, $\omega^{(C)}$, that is coupled to a two-level atomic transition $\omega^{(A)}$;
    see Fig. \ref{fig:LeakyCavity}.
    The detuning is given by the difference of both frequencies. 
    The included imperfections are as follows:
        the cavity mode is damped by losses through partially transmitting cavity mirrors;
        in addition to wanted in- and out-coupling, the atom can spontaneously emit out of the side of the cavity;
        and photons may be absorbed and scattered by cavity mirrors \cite{DiFidio}.

    To describe the open quantum dynamics \cite{Dalibard,Dum,Carmichael_4} of the system under study, we can apply the following master equation for the density operator $\hat \rho(t)$ of the atom-cavity system:
    \begin{align}
        \label{eq:master_0}
        \frac{d \hat \rho(t)}{d t}
        =&{}
        \frac{1}{i\hbar} \big[ \hat H, \hat \rho(t)\big]
        \\ \nonumber
        &{}
        +\Gamma^{(C)} \left(  \hat a \hat \rho(t) \hat a^\dagger -\frac{1}{2}\big\{\hat a^\dagger \hat a, \hat \rho(t)\big\} \right)
        \\ \nonumber
        &{}
        +\Gamma^{(A)}\left( \hat A^{(0,1)} \hat \rho(t) \hat A^{(1,0)} -\frac{1}{2}\big\{ \hat A^{(1,1)}, \hat \rho(t)\big\} \right).
    \end{align}
    Therein, $\hat a$ and $\hat a^\dagger$ are annihilation and creation operators for the cavity field, respectively.
    For the two-level atom, we define the operators $\hat A^{(i,j)}=|i\rangle\langle j|$, with $i,j\in\{0,1\}$.
    The rate $\Gamma^{(C)}$ includes all optical losses, and $\Gamma^{(A)}$ is the spontaneous emission rate of the atom.

    The Hamiltonian $\hat H$ in Eq. \eqref{eq:master_0} describes the free evolution as well as the atom-cavity interaction,
    \begin{equation}
        \label{hamiltonian_0}
        \frac{\hat H}{\hbar}
        = \omega^{(C)} \hat a^\dagger \hat a
        + \omega^{(A)} \hat A^{(1,1)} 
        + g\hat a \hat A^{(1,0)}
        + g^\ast\hat a^\dagger \hat A^{(0,1)},
    \end{equation}
    with $g$ denoting the atom-cavity coupling constant.
    Note that an intercavity interaction, coupling the incoming and outgoing fields, is not present here but is introduced in the next section.
    It is also worth recalling that solving the master equation, as commonly written in the Lindblad form \cite{Lindblad}, can be achieved through the quantum trajectory approach \cite{Dalibard,Dum,Carmichael_4}.

\section{Cascaded Network Dynamics}
\label{sec:CascadedNetworks}

    With a single node of a network explained in the previous section, we can proceed with connecting those building blocks.
    To this end, we analyze the dynamics of a cascaded network consisting of multiple cavities.
    For simplicity, we assume that all cavities are identical.

\subsection{Full multipartite system}

\paragraph*{Equations of motion.}

    For formulating the dynamics of the full open quantum system, the corresponding master equation in Lindblad form reads \cite{Carmichael_4}
    \begin{equation}
        \label{eq:master}
    \begin{aligned}
        \frac{d \hat \rho(t)}{d t}
        =&{}
        \frac{1}{i\hbar}\left[ \hat H, \hat \rho(t)\right]
        \\
        &{}+\sum_{\hat J \in \mathcal J}\left( \hat J\hat \rho(t) \hat J^\dagger - \frac{1}{2} \left\{ \hat J^\dagger \hat J , \hat \rho(t) \right\} \right) ,
    \end{aligned}
    \end{equation}
    where $\mathcal J$ denotes the set of all jump operators (also known as Lindblad operators).
    The full Hamiltonian is given by
    \begin{equation}
        \label{eq:Hamiltonian_tot}
    \begin{aligned}
        \hat H = \sum_{k} \hat H_k + \hat H_\mathrm{int}. 
    \end{aligned}
    \end{equation}
    Akin to Eq. \eqref{hamiltonian_0}, $\hat H_k$ here describes the atom-cavity interaction of each network node and is similarly given by
    \begin{equation}
        \label{eq:hamiltonian_k}
        \frac{\hat H_k}{\hbar}
        = \omega^{(C)} \hat a_k^\dagger \hat a_k
        {+} \omega^{(A)}\hat A_k^{(1,1)}
        {+} g\hat a_k \hat A_k^{(1,0)}
        {+} g^\ast\hat a_k^\dag \hat A_k^{(0,1)},
    \end{equation}
    where the subscript $k$ indexes the cavity mode and the atom at a specific node.
    Moreover, the interaction Hamiltonian $\hat H_\mathrm{int}$ describes the coupling between the various quantized cavity modes,
    \begin{equation}
        \label{eq:hamiltonian_int}
      \hat H_{\rm int}= \hbar \sum_{l,m} \kappa_{l,m} \hat a_l^\dagger \hat a_m ,
    \end{equation}
    with the coupling $\kappa_{l,m}$ between cavities $l$ and $m$, obeying the conditions $\kappa_{l,m}=\kappa_{m,l}^\ast$ and $\kappa_{l,l}=0$ (no self-coupling).
    For coupling multiple cavities, other than in Fig. \ref{fig:LeakyCavity}, one can, for example, utilize a ring-cavity configuration with multiple mirrors, each of which provides additional input and output ports for coupling \cite{Kruse,Cox}.

    The jump operators in Eq. \eqref{eq:master} describe optical losses and spontaneous emission by the two-level atom in each cavity subsystem $k$.
    These operators read
    \begin{equation}
        \label{eq:Jump_1_2}
        \hat J_k^{(C)} = \sqrt{\Gamma^{(C)}} \hat a_k
        \quad\text{and}\quad
        \hat J_k^{(A)} = \sqrt{\Gamma^{(A)}} \hat A_k^{(0,1)},
    \end{equation}
    where $\Gamma^{(C)}$ is the optical loss rate and $\Gamma^{(A)}$ is the spontaneous emission rate.
    Without loss of generality, we chose real-valued and non-negative rates, $\Gamma^{(C)}, \Gamma^{(A)} \in \mathbb{R}_{\geq 0}$.

\paragraph*{State description.}

    As a proof of concept, we restrict ourselves to a single excitation (likewise, one quantum walker) throughout this work.
    For the purpose of quantum walks, the $k$th mode acts as the position-like degree of freedom, and the state of the two-level atom functions as the quantum coin.
    For convenience, we further define the following state vectors:
    \begin{equation}
        \label{eq:ground}
    \begin{aligned}
        &{}
        |g\rangle = \bigotimes_{k} (|0\rangle  \otimes |0\rangle) =  \bigotimes_{k}  |0_k, 0_k\rangle,
        \\
        &{}
        |1_k^{(C)}\rangle = \hat a_k^\dagger |g\rangle,
        \quad \text{and} \quad
        |1_k^{(A)}\rangle =  \hat A_k^{(1,0)} |g\rangle.
    \end{aligned}
    \end{equation}
    Here, $|g\rangle$ is the state of the network in which all the atoms are found in their ground state and all the cavity modes are void of photons, i.e., avacuum state.
    In addition, $|1_k^{(C)}\rangle$ says that the $k$th cavity is occupied with one photon, and $| 1_k^{(A)}\rangle$ indicates that the $k$th two-level atom is excited.

    In order to evaluate the evolution, one can utilize a quantum trajectory approach \cite{Dalibard,Dum,Carmichael_4}.
    That is, the evolution of the system is governed by a nonunitary Schr\"odinger equation that evolves the normalized initial state $|\bar\psi(t_0)\rangle $ into an unnormalized state vector $|\bar \psi (t)\rangle$, as explained in more details below.
    Furthermore, this linear evolution can be randomly interrupted by one of the jumps $\hat J$ in Eq. \eqref{eq:Jump_1_2}.
    Once a jump has occurred at time $t_J$, the wave vector is found to be collapsed into the state $|g\rangle$ because of the action of the jump operator, $\hat J |\bar \psi (t_J)\rangle \mapsto |g\rangle$.
    Thus, for the problem under study, we may have one jump at most.
    After the collapse, the state $|g\rangle$ remains unchanged.
    Therefore, the density operator $\hat \rho(t)$ is an ensemble average over the different trajectories, yielding the incoherent mixture
    \begin{equation}
        \label{eq:rho_t}
        \hat \rho(t)
        = |\bar \psi (t) \rangle \langle \bar \psi (t)|
        + \left(
            1-\langle \bar \psi (t)|\bar \psi (t) \rangle
        \right) |g\rangle\langle g|.
    \end{equation}

\subsection{Analytic solution}

    For the linear (no jump) contribution of the evolution, the system's initial, normalized state is $|\bar\psi(t_0) \rangle $.
    To determine the state vector of the system at a later time $t>t_0$, we solve the nonunitary Schr\"odinger equation \cite{Dalibard,Dum,Carmichael_4}
    \begin{equation}
        \label{eq:Schroedinger_nU}
        i\hbar \frac{d}{d t} |\bar \psi (t) \rangle = \hat H' | \bar \psi (t) \rangle,  
    \end{equation}
    where $\hat H'$ is the non-Hermitian Hamiltonian that takes the form
    \begin{equation}
        \label{eq:Hamiltoinian_n_H}
        \hat H' = \hat H - \frac{i\hbar}{2}\sum_{\hat J \in \mathcal J} \hat J^\dagger \hat J. 
    \end{equation}
    Thus, the system evolves via Eq. \eqref{eq:Schroedinger_nU} into the unnormalized state
    \begin{equation}
        \label{StateVector_nN}
    \begin{aligned}
        |\bar \psi (t) \rangle 
        ={}&
        \sum_{k} \left( \alpha_k(t) |1_k^{(C)}\rangle + \beta_k(t)|1_k^{(A)}\rangle \right)
        \\
        ={}&
        |\vec \alpha(t),\vec 0\rangle
        + |\vec 0,\vec \beta(t)\rangle,
    \end{aligned}
    \end{equation}
    where we combined cavity and atomic probability amplitudes for one excitation in the vectors $\vec \alpha(t)=[\alpha_k(t)]_k$ and $\vec \beta(t)=[\beta_k(t)]_k$, respectively, and $\vec 0=[0]_k$.

\begin{figure*}
	\includegraphics[width=\textwidth]{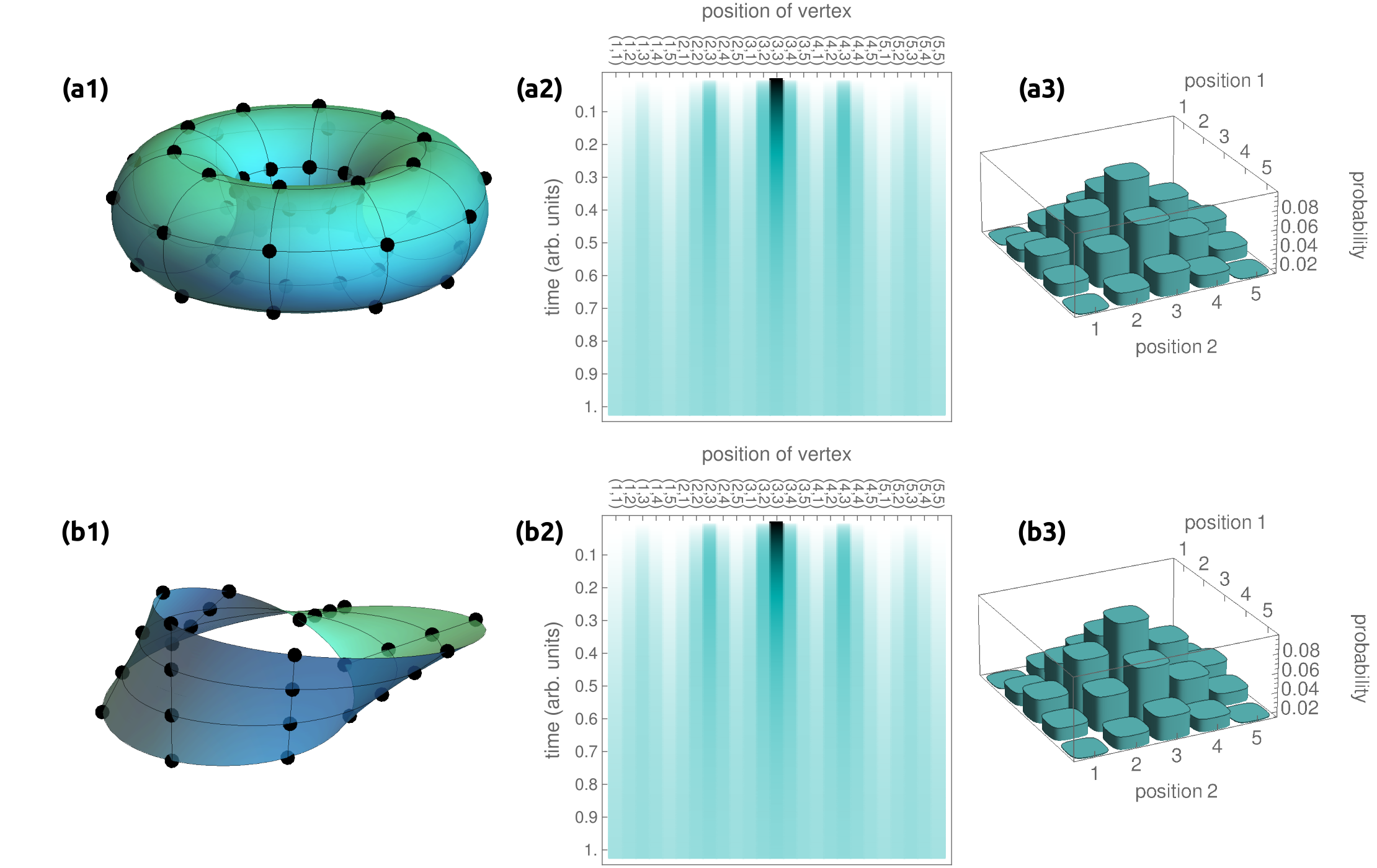}
	\caption{%
        Random walks on graphs.
        We consider different, interesting geometries, i.e., (a1) a torus as a nontrivial topology and (b1) a M\"obius strip as a nonorientable manifold, where vertices are depicted as black circles and connecting edges as black lines.
        Vertices are located as pairs of positions across and around the torus and M\"obius strip.
        For later comparisons with the quantum walks, classical random walks are implemented in (a2) and (b2) for both graphs \cite{comment:RandomWalk}.
        The walker is initialized at vertex $(3,3)$ and randomly proceeds to neighboring positions. 
        With propagation in time, we see a diffusive behavior, eventually leading to a uniform distribution.
        Here we made the particular choice of 25 vertices, i.e., $N_1=N_2=5$.
        In panels (a3) and (b3), the probability distribution of an intermediate time is shown for the two geometries.
        For the quantum walk, cavities resemble the vertices of the graphs and the coupling between cavities is indicated by edges.
	}\label{fig:GraphRandomWalk}
\end{figure*}

    The non-Hermitian Hamiltonian in Eq. \eqref{eq:Hamiltoinian_n_H} acts on the single-excitation states $|1_k^{(C)}\rangle$ and $|1_k^{(A)}\rangle$ as follows:
    \begin{equation}
        \label{eq:Hamiltoinian_n_H_1}
    \begin{aligned}
        \hat H' |1_k^{(C)}\rangle
        ={}& \left(\hbar \omega^{(C)} -\frac{i \hbar}{2}\Gamma^{(C)}\right)|1_k^{(C)}\rangle + \hbar g |1_k^{(A)}\rangle
        \\
        {}&+ \hbar \sum_{j} \kappa_{j,k}|1_j^{(C)}\rangle,
        \\
        \hat H' |1_k^{(A)}\rangle
        ={}&
        \left(\hbar \omega^{(A)} -\frac{i \hbar}{2}\Gamma^{(A)}\right)|1_k^{(A)}\rangle + \hbar g^\ast |1_k^{(C)}\rangle.
    \end{aligned}
    \end{equation}
    From Eqs. \eqref{eq:Schroedinger_nU}, \eqref{StateVector_nN}, and \eqref{eq:Hamiltoinian_n_H_1}, we thus deduce the following equation of motion for the probability amplitudes:
    \begin{align}
        \label{eq:system}
        i \frac{d}{d t}
        \begin{bmatrix}
            {\vec \alpha } (t)
            \\
            {\vec \beta } (t)
        \end{bmatrix} 
        = {}&
        \boldsymbol\eta
        \begin{bmatrix}
            {\vec \alpha } (t)
            \\
            {\vec \beta } (t)
        \end{bmatrix},
        \quad\text{with}
        \\ \nonumber
        \boldsymbol\eta
        = &
        \begin{bmatrix}
            \left(\omega^{(C)} {-}\frac{i}{2}\Gamma^{(C)}\right) \boldsymbol{\mathbbm 1} {+} \boldsymbol K & g^\ast \boldsymbol{\mathbbm 1}
            \\
            g \boldsymbol{\mathbbm 1} & \left(\omega^{(A)} {-}\frac{i}{2}\Gamma^{(A)}\right) \boldsymbol{\mathbbm 1}
        \end{bmatrix},
    \end{align}
    which includes the identity matrix $\boldsymbol{\mathbbm 1}=[\delta_{i,j}]_{i,j}$ and the coupling matrix $\boldsymbol K=[\kappa_{i,j}]_{i,j}=\boldsymbol K^\dag$.
    This yields the solution
    \begin{equation}
       \label{eq:system_solutions}
        \begin{bmatrix}
            {\vec \alpha } (t)
            \\
            {\vec \beta } (t)
        \end{bmatrix} 
        = \exp{(-it\boldsymbol\eta)}
        \begin{bmatrix}
            {\vec \alpha } (t_0)
            \\
            {\vec \beta } (t_0)
        \end{bmatrix}.
    \end{equation}
    More explicitly, the propagator $\exp{(-it\boldsymbol\eta)}$ can be decomposed as follows:
    \begin{align}
        \label{eq:solution_exp}
        {}& \exp{(-it\boldsymbol\eta)}
        \\ \nonumber
        ={}& 
        \begin{bmatrix}
            \boldsymbol E & 0 \\ 0 & \boldsymbol E
        \end{bmatrix}
        \left(
            \begin{bmatrix}
                \cos(t\boldsymbol\Omega) & 0 \\ 0 & \cos(t\boldsymbol\Omega)
            \end{bmatrix}
        \right.
        \\ \nonumber
        &{}
        \left.
            -i\begin{bmatrix}
                \boldsymbol\Omega^{-1} & 0 \\ 0 & \boldsymbol\Omega^{-1}
            \end{bmatrix}
            \begin{bmatrix}
                \sin(t\boldsymbol\Omega) & 0 \\ 0 & \sin(t\boldsymbol\Omega)
            \end{bmatrix}
            \begin{bmatrix}
                \boldsymbol\Delta & g^\ast\boldsymbol{\mathbbm 1}
                \\
                g\boldsymbol{\mathbbm 1} & -\boldsymbol\Delta
            \end{bmatrix}
        \right),
    \end{align}
    which itself contains the matrices defined via
    \begin{equation}
        \label{eq:symbols}
    \begin{aligned}
        2\boldsymbol\Delta
        ={}&
        \bar\omega \boldsymbol{\mathbbm 1} + \boldsymbol K,
        \quad
        \boldsymbol\Omega^2
        =
        \boldsymbol\Delta^2+ |g|^2 \boldsymbol{\mathbbm 1},
        \\
        \text{and}\quad
        \boldsymbol E
        ={}& e^{-it\bar\nu/2}
        \exp\left(-\frac{it}{2}\boldsymbol K\right),
    \end{aligned}
    \end{equation}
    where $\bar\omega=\omega^{(C)} -i\Gamma^{(C)}/2-\omega^{(A)} + i\Gamma^{(A)}/2$ and $\bar\nu=\omega^{(C)} -i\Gamma^{(C)}/2+\omega^{(A)} - i\Gamma^{(A)}/2$.
    Note that all matrices commute with the coupling matrix $\boldsymbol K$, thus sharing a joint diagonalization.

    The above solution is applied to our computations leading to the results presented in the remainder of this work.
    In particular, we use the time-step iteration $[\vec \alpha(t+dt),\vec \beta(t+dt)]^\mathrm{T}=\boldsymbol S[\vec \alpha(t),\vec \beta(t)]^\mathrm{T}$, with $\boldsymbol S=\exp{(-i\,dt\,\boldsymbol\eta)}$, to propagate probability amplitudes over a small-time differential, $dt$.

\section{Continuous-in-time quantum walks on different geometries}
\label{sec:Geometries}

    In this section, we study different network geometries by analyzing how the initially localized excitation propagates through the network.
    By exploring specific examples of a nontrivial quantum-walk dynamics, we demonstrate the flexibility of the model.
    The two cases under study pertain to a network on a closed surface with a toroidal structure and a network on a two-dimensional nonorientable surface in a M\"obius-strip configuration; 
    see, also, Fig. \ref{fig:GraphRandomWalk}, including the classical random-walk evolution.

\subsection{Considered scenarios and random walk}

    The characterization of the chosen geometry is realized by specifying the particular interaction Hamiltonian, given by Eq. \eqref{eq:hamiltonian_int}, that describes the interaction between the different cavities.
    This requires one to specify the matrix $\boldsymbol K=[\kappa_{i,j}]_{i,j}$ in  Eq. \eqref{eq:system}.
    To investigate different geometries, we consider a network as a graph wherein the vertices, i.e., nodes, are the cavities together with the two-level atom, and edges of the graph define the network connectivity.
    By applying Eq. \eqref{eq:system_solutions}, the dynamics of the system in the chosen network geometry is then fully determined.

    In the scenarios described above, the adjacency matrix $\boldsymbol{A}$, well known from graph theory \cite{Nica}, is proportional to the coupling matrix.
    Further, we assume a uniform $\kappa$ coupling, where $\kappa>0$, between connected vertices and undirected channels, $\boldsymbol A=\boldsymbol A^\mathrm{T}$.
    Therefore, we can write $\boldsymbol K$ as
    \begin{equation}
        \label{eq:kappa_matrix}
        \boldsymbol K= \kappa \boldsymbol{A}=\kappa[A_{i,j}]_{i,j},
    \end{equation}
    where $A_{i,j}=1$ if the vertices $i$ and $j$ share an edge and zero otherwise.
    While not considered here, we could also model non-uniform couplings between nodes (i.e., weighted graphs) and one-way network connections (i.e., directed graphs) through values of the adjacency matrix other than zero and one and $\boldsymbol A\neq \boldsymbol A^\mathrm{T}$, respectively.

\paragraph*{Torus.}

    The first geometry we consider is a closed surface with a toroidal structure; see Fig. \ref{fig:GraphRandomWalk}(a1).
    Topologically speaking, the torus has a nontrivial genus of one.
    A graph on a torus is characterized by nodes $i=(i_1,i_2)$, and a two-dimensional grid with $1 \leq i_1,j_1\leq N_1$ and $1 \leq i_2,j_2\leq N_2$.
    In addition, a periodic boundary condition is separately imposed on the nodes in both components.
    Therefore, the entries of the adjacency matrix $\boldsymbol{A}$ can be expressed as
    \begin{equation}
        \label{eq:adjacency_matrix_torus}
    \begin{aligned}
        {}&A_{i,j}
        =
        A_{(i_1,i_2),(j_1,j_2)}
        \\
        ={}&
        \delta_{i_1,j_1+1 \mathop{\mathrm{mod}} N_1}\delta_{i_2,j_2}
        +\delta_{i_1,j_1-1 \mathop{\mathrm{mod}} N_1}\delta_{i_2,j_2}
        \\
        {}&
        +\delta_{i_1,j_1}\delta_{i_2,j_2+1 \mathop{\mathrm{mod}} N_2}
        +\delta_{i_1,j_1}\delta_{i_2,j_2-1 \mathop{\mathrm{mod}} N_2}.
    \end{aligned}
    \end{equation}

    In Fig. \ref{fig:GraphRandomWalk}(a2), we show the classical random walk on this graph, using a common rate equation \cite{comment:RandomWalk}.
    Starting with a determined position of the walker, neighboring nodes become excited, spreading across and around the torus. 
    As one would expect, the classical evolution is a dispersion process that converges to the uniform probability distribution of the walker on the graph.

\paragraph*{M\"obius strip.}

    The second example pertains to a two-dimensional graph configuration on a M\"obius strip; cf. Fig. \ref{fig:GraphRandomWalk}(b1).
    In contrast to the previous graph, this strip corresponds to a nonorientable surface.
    For this graph, the adjacency matrix $\boldsymbol A$ is determined by the entries
    \begin{equation}
        \label{eq:adjacency_matrix_moebius}
    \begin{aligned}
        A_{(i_1,i_2),(j_1,j_2)}
        ={}&
        \delta_{i_1,j_1+1}\delta_{i_2,j_2}
        +\delta_{i_1,j_1-1}\delta_{i_2,j_2}
        \\
        {}&
        +\delta_{i_1,j_1}\delta_{i_2,j_2+1}
        +\delta_{i_1,j_1}\delta_{i_2,j_2-1}
        \\
        {}&+\delta_{i_1,1}\delta_{j_1,N_1}\delta_{i_2+j_2,N_2+1}
        \\
        {}&+\delta_{i_1,N_1}\delta_{j_1,1}\delta_{i_2+j_2,N_2+1},
    \end{aligned}
    \end{equation}
    with $1 \leq i_1,j_1\leq N_1$ and $1 \leq i_2,j_2\leq N_2$.
    Figure \ref{fig:GraphRandomWalk}(b2) depicts the classical random walk for such a M\"obius-strip configuration.
    The resulting diffusion behaves rather similarly to the previous example of a torus in this classical dynamics, converging to a uniformly spread probability of the walker's position.

\begin{figure*}
	\includegraphics[width=\textwidth]{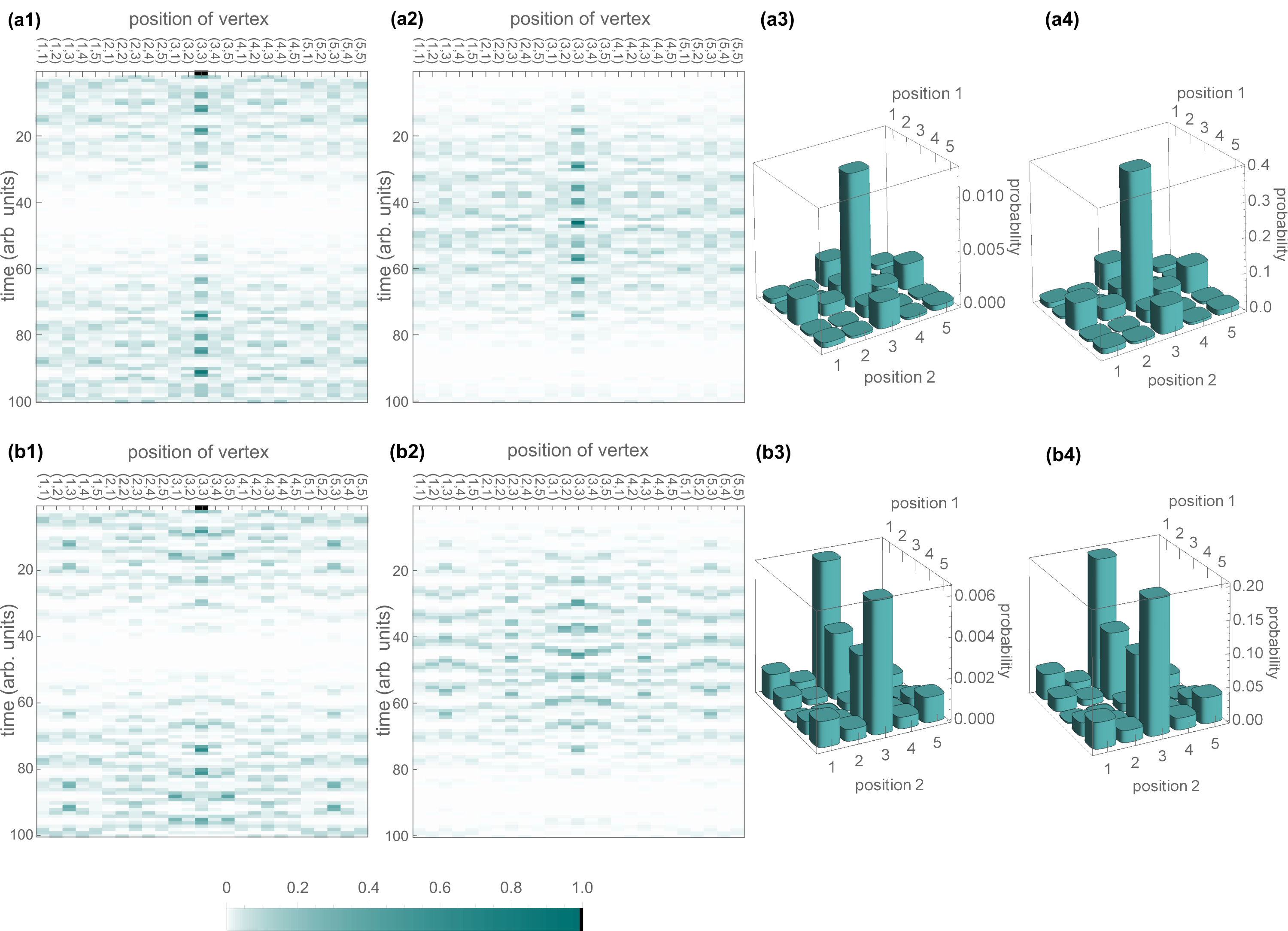}
	\caption{%
        Dynamics of the quantum walk for a network with toroidal structure (top row) and a M\"obius strip (bottom row). 
        The probability of finding the walker at a certain position as a cavity excitation and atom excitation is shown in the first and second columns, respectively. 
        Accordingly, the evolution of the cavity probability $|\alpha_{k=(k_1,k_2)}(t)|^2$ on each configuration is shown in (a1) and (b1), and the evolution of the atomic probability $|\beta_{k=(k_1,k_2)}(t)|^2$ is shown in (a2) and (b2) for discrete time steps $t\to t+dt$.
        In the last two columns, the probability distribution of an intermediate time is depicted.
        In particular, for a chosen time step $t=40\,dt$, the cavity probability $|\alpha_{k=(k_1,k_2)}|^2$ is shown in (a3) and (b3), and the atomic probability $|\beta_{k=(k_1,k_2)}|^2$ is shown in (a4) and (b4).
        For the strong-coupling regime illustrated here, the considered parameters in units of the coupling $\kappa$ are $\Gamma^{(C)}/\kappa=\Gamma^{(A)}/\kappa=0$, $\omega^{(A)}/\kappa=1$, $\omega^{(C)}/\kappa=0.9$, and $g/\kappa=10^5$, while defining $dt=1/\kappa$ in the numerical analysis.
    }\label{fig:Probabilities}
\end{figure*}    

\subsection{Quantum evolution}

    After having introduced the geometries of interest and the dynamics of classical random walks, we now analyze the quantum evolution described by our model.
    For this purpose, we compute the dynamics of a quantum walk according to Eq. \eqref{eq:system_solutions} for both geometries.
    For the sake of comparison, we analyze the same scenario presented in Fig. \ref{fig:GraphRandomWalk};
    that is, a network of $N_1=N_2=5$ cavities and, as the initial condition, a single excitation of the cavity field located at vertex $(3,3)$ and no excitation of atoms, $|\bar{\psi}(t_0=0)\rangle=|1^{(C)}_k\rangle$ with $k=(3,3)$.

    The results of our computation are presented in Fig. \ref{fig:Probabilities}, where the top row corresponds to the torus and the bottom one to the M\"obius strip. 
    The evolution of the probability is given as a timeline in the first two columns.
    The probability of finding the excitation as a photon across the nodes, $|\alpha_k|^2$, appears in Figs. 3(a1) and (b1), whereas the probability of finding an excited atom, $|\beta_k|^2$, is presented in Figs.(a2) and (b2).
    Note that occupation probability significantly alternates between light and matter degrees of freedom with a frequency modulated by $\boldsymbol\Omega$; cf. Eqs. \eqref{eq:solution_exp} and \eqref{eq:symbols}.
    The most striking signature of the quantum behavior is the interference between the probability amplitudes, which gives rise to rich interference patterns in the probability distribution.
    
    This interference, which can already be seen in the four timelines, is even more clear if we inspect one specific time step. This is shown in Figs. \ref{fig:Probabilities}(a3) and \ref{fig:Probabilities}(b3) for photon probabilities and in Figs. \ref{fig:Probabilities}(a4) and \ref{fig:Probabilities}(b4) for excited-atom probabilities. 
    By comparing the same time step in the torus and M\"obius strip, one can see the difference between the structures of the interference, which is not observable in the random walk analysis in Fig. \ref{fig:GraphRandomWalk}.
    In particular, there is a clear asymmetry between the vertices' position component $1$ and $2$ for the probability distribution in the M\"obius strip.
    By contrast, position components $1$ and $2$ appear interchangeable for the torus in this particular scenario. 
    Importantly, we emphasize that all interference patterns noticeably contrast with the interference-free distribution in the random-walk scenario; 
    compare Figs. \ref{fig:GraphRandomWalk}(a3) and \ref{fig:GraphRandomWalk}(b3) with Figs. \ref{fig:Probabilities}(a3), \ref{fig:Probabilities}(a4), \ref{fig:Probabilities}(b3), and \ref{fig:Probabilities}(b4). 

    For the sought-after comparison with the random walk, we assumed lossless cases here;
    the impact of quantum jumps is discussed in the next section together with characterizing the systems' entanglement.
    Furthermore, one way to extend our model is to introduce an optical intercavity field that couples to the internal field of connected cavities studied here. 
    Losses of intercavity fields act as an additional decoherence mechanism, leading to a transition from strong interference to classical diffusion as observed in a random walk \cite{Paris2,Karski,DiFidio_1}.

\section{Multipartite entanglement evolution}
\label{sec:EntanglementEvolution}

    In the previous section, we focused on the classical and quantum probabilities to find the walker in a certain configuration.
    However, this does not necessarily prove that quantum correlations are generated through the evolution.
    Therefore, we now provide a detailed entanglement analysis of the systems under consideration.

\subsection{Entanglement witnesses}

    One experiment-friendly way to assess entanglement is given in terms of measurable operators, dubbed entanglement witnesses \cite{Horodecki, Horodecki2, Toth, Terhal}.
    One way to construct such witnesses is based on finding the maximal (likewise, minimal) expectation values of an observable $\hat L$ for separable states via generalized eigenvalue problems \cite{Sperling_0, Sperling_1}, the so-called separability eigenvalue equations.
    The thereby constructed witnesses are directly applicable to detecting multipartite entanglement in experiments \cite{Gutierrez, Gerke_1, Gerke_2}.

\paragraph*{Characterization of entanglement.}

    The separability criterion under study, for an observable $\hat L$, is expressed by the following bound \cite{Toth, Sperling_1}:
    \begin{equation}
        \label{eq:L_general}
        \langle \hat L\rangle
        =\mathrm{tr}[\hat \rho_{\text{sep}} \hat L]
        \leq \sup_{|\psi\rangle\in\mathcal{S}}\{\langle\psi|L|\psi\rangle\}
        =g_{\max},
    \end{equation}
    where $\hat\rho_\text{sep}$ is a nonentangled state and $\mathcal{S}$ denotes the set of all pure separable states.
    We have indicated with $g_{\max}$ the largest possible expected value of $\hat L$ for separable states.
    This maximal expectation value is identical to the maximal separability eigenvalue from the construction approach mentioned before \cite{Sperling_0, Sperling_1}.

    A violation of the inequality in Eq. \eqref{eq:L_general} certifies entanglement.
    As a consequence, an expected value exceeding the discussed maximum can be utilized to quantify the amount of entanglement present in the state,
    \begin{equation}
        \label{eq:EntanglementQuantifier}
        E=\begin{cases}
            0
            &
            \text{if Eq. \eqref{eq:L_general} is obeyed,}
            \\
            \dfrac{\langle \hat{L}\rangle-g_{\max}}{\lambda_{\max}-g_{\max}}
            &
            \text{otherwise,}
        \end{cases}
    \end{equation}
    where $\lambda_{\max}$ denotes the maximal expected value for all separable and inseparable states, i.e., the maximal ordinary eigenvalue of $\hat L$, and is, for all examples considered here, $\lambda_{\max}=1$.
    Thus, $E$ yields a value between zero and one and is zero for separable states, hence quantifying the witnessed entanglement \cite{Eisert}.

\paragraph*{Observables under consideration.}

    For distinguishing quantum correlations, we specifically consider four kinds of entanglement: multipartite entanglement of the full system, multipartite entanglement between the atoms, multipartite entanglement between the cavity modes, and bipartite light-matter entanglement.

    For the former three forms of multipartite entanglement, we take an observable that is based on a generalized, $N$-partite $W$ state with phases
    \begin{equation}
        \label{eq:L_at}  
    \begin{aligned}
        \hat{L}={}&|W\rangle\langle W|,
        \quad\text{with}
        \\
        |W\rangle={}&\dfrac{1}{\sqrt{N}} \sum_{k=1}^N e^{i\varphi_k}
        |0\rangle^{\otimes (k-1)}\otimes|1\rangle\otimes|0\rangle^{\otimes (N-k)},
    \end{aligned}
    \end{equation}
    where the phases $\varphi_k$ are adjusted such that they coincide with the single-excitation state under study for maximizing the expectation value $\langle\hat L\rangle$.
    Regardless of the phases, the maximum expected value for separable states, i.e., maximal separability eigenvalue, reads \cite{Shimoni,Pagel}
    \begin{equation}
        \label{eq:MaxBoundW}
        g_{\max}=\left(\frac{N-1}{N}\right)^{N-1}.    
    \end{equation}
    Recall that $\lambda_{\max}=1$ holds true for this rank-one operator $\hat L$ that is based on a normalized state.

    With the observable described thus far, we probe how similar the system's state is to an $N$-partite, maximally multipartite-entangled state in a $W$ configuration that includes one excitation.
    For the states produced in our quantum walk scenario, i.e., Eqs. \eqref{eq:rho_t} and \eqref{StateVector_nN}, we can directly compute the sought-after expectation values.
    Since we have $N=N_1N_2$ atoms and cavities for the considered geometries, for the cavity-only entanglement, we find
    \begin{equation}
        \langle\hat L\rangle=\frac{1}{N}\left(\sum_{k}|\alpha_k|\right)^2,
    \end{equation}
    where the phases of the $N$-partite $W$ state are $\varphi_k=\arg\alpha_k$.
    Analogously, the entanglement between atoms is determined through the expectation value
    \begin{equation}
        \langle\hat L\rangle=\frac{1}{N}\left(\sum_{k}|\beta_k|\right)^2.
    \end{equation}
    When considering the entire system consisting of $N$ cavities and $N$ atoms together, we have a $2N$-partite $W$ state and
    \begin{equation}
        \langle\hat L\rangle=\frac{1}{2N}\left[
            \sum_k \left(
                |\alpha_k|+|\beta_k|
            \right)
        \right]^2.
    \end{equation}
    Note that the separable bound in Eq. \eqref{eq:MaxBoundW} here straightforwardly generalizes to $g_{\max}=([2N-1]/[2N])^{2N-1}$.

    For the bipartite light-matter entanglement, we similarly consider an observable $\hat L=|\Phi\rangle\langle\Phi|$, but here with a normalized state $|\Phi\rangle$ parallel to the unnormalized system state in Eq. \eqref{StateVector_nN},
    \begin{equation}
        |\Phi\rangle=\frac{|\vec\alpha,\vec 0\rangle+|\vec0,\vec\beta\rangle}{\sqrt{\langle\vec \alpha|\vec \alpha\rangle+\langle\vec\beta|\vec \beta\rangle}},
    \end{equation}
    also yielding $\lambda_{\max}=1$.
    Note that $|\vec 0\rangle$ is free of excitations and, thus, perpendicular to the single-excitation cavity state $|\vec\alpha\rangle$ and the single-excitation atomic state $|\vec\beta\rangle$.
    As $|\Phi\rangle$ is thus given in its Schmidt decomposition, the bipartite cavity-atom separability bound $g_{\max}$ can be readily found via the maximal Schmidt coefficient \cite{Sperling_0},
    \begin{equation}
        g_{\max}=\frac{\max\left\{
            \langle\vec\alpha|\vec\alpha\rangle,
            \langle\vec\beta|\vec\beta\rangle
        \right\}}{\langle\vec \alpha|\vec \alpha\rangle+\langle\vec\beta|\vec \beta\rangle}.
    \end{equation}
    Also, the expectation value can be readily determined as $
        \langle\hat L\rangle
        =\langle\Phi|\hat\rho|\Phi\rangle
        =\langle\vec \alpha|\vec \alpha\rangle+\langle\vec\beta|\vec \beta\rangle
    $, using Eqs. \eqref{eq:rho_t} and \eqref{StateVector_nN}.

\begin{figure*}[t]
	\includegraphics[width=0.98\textwidth]{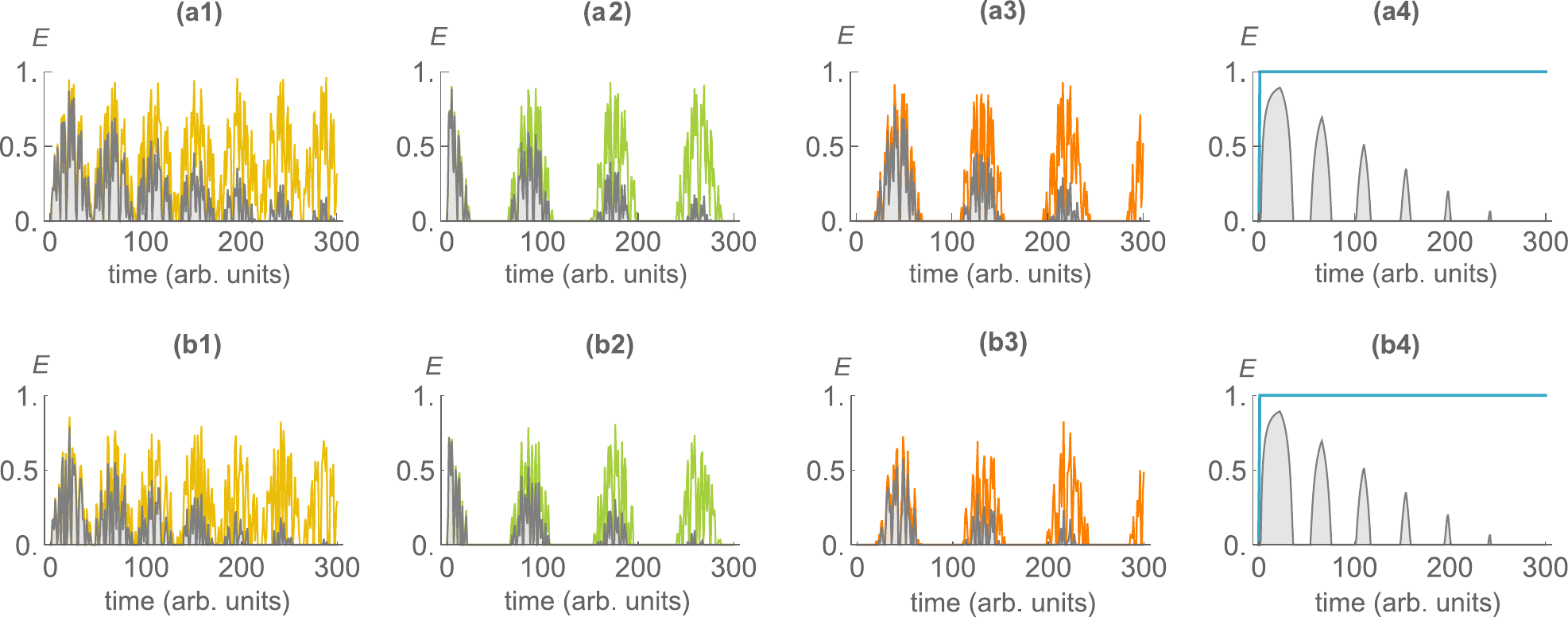}
	\caption{%
        Evolution of the entanglement in the quantum walk for a network with toroidal structure (top row) and a M\"obius strip (bottom row) in terms of the entanglement quantifier in Eq. \eqref{eq:EntanglementQuantifier}. 
        Lines (in bright color) represent the ideal, lossless entanglement evolution ($\Gamma^{(C)}/\kappa=\Gamma^{(A)}/\kappa=0$) whereas decaying curves (in gray) show attenuated situations ($\Gamma^{(C)}/\kappa=\Gamma^{(A)}/\kappa=0.0025$).
        The multipartite entanglement of the full system is shown in (a1) and (b1) for the two geometries.
        The multipartite entanglement between the cavity modes alone is shown in (a2) and (b2), and the analogous case for atoms in (a3) and (b3). 
        In (a4) and (b4), the bipartite light-matter entanglement is shown.
        The other parameters---in addition to the decay rates---are the same as in Fig. \ref{fig:Probabilities}.
    }\label{fig:Entanglement}
\end{figure*}

\subsection{Entanglement evolution}

    The evolution of the aforementioned four kinds of entanglement is displayed in Fig. \ref{fig:Entanglement} for the torus (top row) and M\"obius strip (bottom row).
    In the four depicted scenarios, using $E$ as given in Eq. \eqref{eq:EntanglementQuantifier}, we consider both a lossless evolution (brighter colors) and an evolution subject to attenuations (gray).
    The multipartite entanglement of the full system is shown in the first column.
    This includes all quantum correlations between the $2N$ elements (all cavity modes and atoms) and is $E=1$ for a maximal entanglement in a $W$-state configuration of the single excitation. 
    The second column similarly depicts the entanglement between all cavity fields alone, being the $W$ entanglement of the system after tracing over the atomic degrees of freedom.
    Likewise, the entanglement between the atoms is shown in the third column.
    Finally, the bipartite light-matter entanglement, i.e., the entanglement between the collection of all atoms and the collection of all cavity fields, is presented in Figs.  \ref{fig:Entanglement}(a4) and \ref{fig:Entanglement}(b4).

    When there are no optical losses and no spontaneous decays in the system, we can see in the fourth column of Fig. \ref{fig:Entanglement} how the entanglement between light and matter is constant and maximum, $E(t)=1$, except for the initial time,  $E(t_0=0)=0$, as we prepare the system in the separable configuration $|\bar\psi(t_0)\rangle=|1_k^{(C)}\rangle$, with $k=(3,3)$. 
    This discontinuous behavior reflects the fact that as soon as the evolution of the initial photon excitation gives a non-zero probability of finding an atomic excitation, the light and matter subsystems become entangled.
    When losses are included, however, these perfect correlations decay, as seen in Figs. \ref{fig:Entanglement}(a4) and \ref{fig:Entanglement}(b4).
    
    The multipartite entanglement of the full system, shown in Figs. \ref{fig:Entanglement}(a1) and \ref{fig:Entanglement}(b1), is the most robust kind of entanglement against losses since it is the one with the highest degree of separation, i.e., each atom and cavity mode is considered as one subsystem.
    Note that even if concrete values are not identical for both geometries, the general behavior is analogous for the quantum walks on torus and M\"obius-strip graphs, including the time decay due to losses, for all kinds of entanglement.
    Furthermore, we observe several instances of increasing and decreasing entanglement, including so-called sudden deaths and sudden births \cite{Mazzola}.
    In general, for the four kinds of entanglement analyzed here, one can observe that a considerable amount of entanglement is created during the evolution and sustained over time, as described by the envelopes of $E(t)$ and notwithstanding short-time oscillations within.

\section{Conclusions}
\label{sec:Conclusion}

    The performance of a multipartite light-matter network with arbitrary configurations as a quantum walk has been considered.
    The quantum dynamics and the multipartite entanglement have been analyzed, focusing on two highly nontrivial geometries: a torus as a genus-one topology and a M\"obius strip as a nonorientable manifold. 
    Clear signatures of quantumness---distinct from the dispersive random-walk evolution---were observed in our computation, which was based on the exact solutions we derived.
    Also, the impact of optical losses and spontaneous, atomic decays were included in our model.

    Quantum interference patterns in the probability distributions were found for both geometries.
    This included oscillatory features in the optical part and in the atomic part, as well as between them. 
    A comparison of random walks and quantum walks further showed that the quantum interference is highly sensitive to the underlying geometry, while no distinctive features were observable in the classical diffusion.

    Additional quantum properties of the network are confirmed by the entanglement created through the open-system evolution. 
    The analysis of entanglement includes the multipartite entanglement of the full system, the multipartite entanglement between the atoms, the multipartite entanglement between the cavity modes, and the bipartite light-matter hybrid entanglement.
    For all these cases and both concrete examples of geometries, we compared the ideal scenario with situations when optical losses and spontaneous emissions are taken into account. 

    Beyond the specific examples, our approach allows for analyzing quantum interference in other geometries, tailored to different, even continuous-in-time quantum simulation tasks.
    One advantage of using a cascaded configuration is that the evolution itself creates interesting kinds of entanglement from an initially nonentangled state. 
    \textcolor{black}{In addition to the system considered here, arrays of optomechanical systems present a promising platform in which our approach may be valuable for implementing quantum walks \cite{Ludwig,Shahmoon}.}
    Thus, we presented a flexible and versatile model that can be used to characterize scalable light-matter networks for interfacing different physical platforms in quantum technologies.

\begin{acknowledgments}
    C.D.F. would like to thank the TQS group for the kind hospitality at Paderborn University.
    J.S. and LA acknowledge financial support through the Ministry of Culture and Science of the State of North Rhine-Westphalia (Project PhoQC).
\end{acknowledgments}


\begin{thebibliography}{00}
\bibitem{comment:RandomWalk}
    The probability of a random walker on a graph evolves via a rate equation, $p_j(t+dt)=p_j(t)+dt\sum_{k}R_{j,k}p_k(t)$.
    For $j\neq k$, $R_{j,k}\geq0$ denotes the transition rate from vertex $k$ to $j$, which here is a constant, $R_{j,k}=r$ if vertices are connected by an edge and $R_{j,k}=0$ otherwise.
    Additionally, one has $R_{k,k}=-\sum_{j:j\neq k}R_{j,k}$ and $dt\leq \min_k\{-R_{k,k}^{-1}\}$.
\end{thebibliography}
\end{document}